\documentclass[pre,a4paper,oneside,twocolumn]{revtex4}
\usepackage[utf8x]{inputenc}
\usepackage{latexsym}
\usepackage{fancyhdr}
\usepackage{color} 
\usepackage{soul} 
\usepackage{graphicx}  

\usepackage[dvipsnames]{xcolor}

\usepackage[pdfborder={0 0 0}]{hyperref}

\usepackage[normalem]{ulem} 

\begin{document}

\title{On the relation between Transversal and Longitudinal Scaling in Cities}

\author{Fabiano L. Ribeiro\textsuperscript{1,2}, Joao Meirelles\textsuperscript{3}, \\ Vinicius M. Netto\textsuperscript{4}, Camilo Rodrigues Neto\textsuperscript{5}, and Andrea Baronchelli\textsuperscript{2}}

\begin{abstract}
Given that a group of cities follows a scaling law connecting urban population with socio-economic or infrastructural metrics (transversal scaling), should we expect that each city would follow the same behavior over time (longitudinal scaling)?
This assumption has important policy implications, although rigorous empirical tests have been so far hindered by the lack of suitable data. 
Here, we advance the debate by looking into the temporal evolution of the scaling laws for 5507 municipalities in Brazil.
We focus on the relationship between population size and two urban variables, GDP and water network length, analyzing the time evolution of the system of cities as well as their individual trajectory.
We find that longitudinal (individual) scaling exponents are city-specific, but they are distributed around an average value that approaches to the transversal scaling exponent when the data are decomposed to eliminate external factors, and when
we only consider cities with a sufficiently large growth rate. Such results give support to the idea that the longitudinal dynamics is a micro-scaling version of the transversal dynamics of the entire urban system. Finally, we propose a mathematical framework that connects the microscopic level to global behavior, and, in all analyzed cases, we find good agreement between theoretical prediction and empirical evidence. 

\end{abstract}

\maketitle

\noindent \textbf{1} Department of Physics (DFI), Federal University of Lavras (UFLA), Lavras MG, Brazil;
\\
\textbf{2} Department of Mathematics, City, University of London,  UK;
\\
\textbf{3}  Department of Civil and Environmental Engineering, Swiss Federal Institute of Technology Lausanne, Lausanne, VD, Switzerland;
\\
\textbf{4}  Universidade Federal Fluminense (UFF), Niterói, RJ, Brasil;
\\
\textbf{5} 
School of Arts, Sciences and Humanities, University of Sao Paulo, Sao Paulo, SP, Brazil.
\\

\section{Introduction}

An unprecedented abundance of data has significantly advanced our understanding of urban phenomena over the past few years \cite{Batty2012,west2017scale, Bettencourt2013_bigdata,barthelemy2016structure}. 
These advances were also enabled by the work of many theorists from different areas, such as physicists, urbanists and complex systems scientists, among others,  
who brought new insights and theories to the field, resulting in a significant step towards a new science of cities \cite{Batty2013}.

A crucial finding concerns the scaling properties of urban systems. 
Empirical evidence has shown that an urban variable,  $Y$,  scales with the population size $N$ of a city, obeying a power law of the kind $Y \propto  N^{\beta}$, where $\beta$ is the scaling exponent quantifying how the urban metric reacts to the population increase \cite{Alves2013, bettencourt2016urban, bettencourt2007growth, Strano2016a, meirelles2018evolution, gomez2012statistics, Kuhnert2006}.
On the one hand, the data revealed that socioeconomic urban variables such as the number of patents, wages, and GDP present a \textit{superlinear} behavior in relation to the population size ($\beta >1$). 
Using the language of economics, one might say that this kind of urban variables exhibits \textit{increasing returns} to the urban scale. 
On the other hand, infrastructure variables such as the number of gas stations and length of roads scale \textit{sublinearly} with the population size ($\beta <1$). 
Finally, there is a third class of variables related to individual basic services, such as household electrical and water consumption, and total employment, which scales linearly with population size ($\beta \approx 1$).

Among the various attempts to explain such behavior in urban phenomena \cite{gomez2017explaining, pumain2006, Ribeiro2017}, one of the most successful was proposed by Bettencourt and colleagues \cite{Bettencourt2013}. 
This theory proposes that urban scaling is a result of an interplay between urban density and diversity, which are related to economic competition and knowledge exchange, respectively. 
This is a consequence of the interaction between the individuals that compose a city, resulting in innovation, economic growth, and economies of scale.

As these scaling laws have been observed in different countries \cite{bettencourt2007growth, bettencourt2016urban, van2016urban, sahasranaman2019urban, Alves2013, adhikari2017growth, Louf2014, meirelles2018evolution, Gomez-Lievano2012, Gomez-Lievano2016} and periods of time \cite{ortman2014pre, cesaretti2016population}, some works also claimed that such patterns are, in fact, the manifestation of a universal law that would generally govern cities regardless of their context, culture, geography, level of technology, policies or history \cite{Bettencourt2013, Louf2014,ribeirocity2017, Gomez-Lievano2016, Gomez-Lievano2012, Strano2016a}. The universality proposition has been challenged \cite{arcaute2015constructing, Cottineau2017, cottineau2015paradoxical}, but most evidence seem to confirm the generality, while exceptions are normally explained by local particularities \cite{strano2016rich,meirelles2018evolution, sahasranaman2019urban,muller2017does}.
According to this proposition, in the long term, the general performance of a particular city would be greatly independent of individual - political - choices: the total amount of social interactions between its citizens would guide, to a great extent, the city towards the observed scaling behavior. 
This proposition is unprecedented in urban science and the identification and validation of such universal dynamics
could help urban policymakers to identify opportunities and improve the life quality of dwellers.

A key open question is the difference in the scaling properties of \textit{single} cities and \textit{systems} of cities. Does an individual city growing in time follow the same scaling pattern observed for a snapshot of a group of cities? 
In the last years, few works have accurately focused on the dynamics of individual cities \cite{Keuschnigg2019, Barthelemy2017, Hong2018, Zhao2018}, while a growing literature has been concentrating on the scaling properties of a set (system) of cities. We call the former \textit{longitudinal} scaling properties, which take into account the evolution of individual cities in time, and the latter \textit{transversal} scaling across an urban system, i.e., computed from the set of cities that compose the system.
Some recent works addressed this issue, reaching no unanimous conclusion. For example, Depersin and Barthelemy analyzed the scaling exponent in time delays in traffic congestion in 101 US cities and found longitudinal scaling to be path-dependent on the individual evolution of cities and unrelated to the transversal scaling, challenging the universality proposition \cite{Barthelemy2017}. In turn, Hong et al. argued that longitudinal and transversal exponents are correlated, but it is essential to eliminate global effects and properly measure the longitudinal scaling exponent \cite{Hong2018}. Another work has found that the power-law scaling of 32 major cities in China could adequately be characterized for both transversal and longitudinal scaling \cite{Zhao2018}. More recent work also analyzed such an issue for the wage income in Sweeden and found superlinear scaling for both longitudinal and transversal scaling, but the former was characterized by larger scaling exponents  \cite{Keuschnigg2019}.

Here, we will present our analysis of the transversal and longitudinal behavior of GDP and water network length (socio-economic and
infrastructure variables, respectively) for 5507 Brazilian municipalities.
Our main results show that the longitudinal scaling exponents are different from each other, as suggested by
Depersin and Barthelemy's work \cite{Barthelemy2017}, but they are distributed around an average that approaches the transversal scaling exponent when the data are decomposed to eliminate external factors and when we consider only subsets of cities with a sufficiently large growth rate. Such results give support to the idea that the longitudinal dynamics is a micro-scaling version of the transversal dynamics of the entire urban system.

The paper is organized as follows: having posed the research problem in this section, we shall unfold our method and data used to assess the evolution of two different urban metrics in section~(\ref{sec_empirical}), namely GDP and water network length as a function of population size in different periods of time (from 1998 to 2014) for 5507 municipalities in Brazil. Section~(\ref{sec_theoretical}) brings details of the theoretical approach used to describe the dynamics of such properties as an analogous problem of particles in a vector field, applied in a way to render the relation between longitudinal and transversal scaling exponents clearer. In this section, we also explore the implications of our findings, along with potential contributions. Finally, we draw our conclusions in  section~(\ref{sec_conclusion}).

\section{Empirical evidence}\label{sec_empirical}

\subsection*{Data of the Brazilian Urban System and its Scaling properties}

The data presented here refer to 5507 Brazilian municipalities, with contiguous dense surrounding areas aggregated in single spatial units from the totality of 5570 Brazilian administrative divisions. Data were collected from the website of the \textit{Brazilian Institute of Geography and Statistics} (IBGE)\footnote{Instituto Brasileiro de Geografia e Estatística, https://www.ibge.gov.br/} and from the water-sewage-waste companies national survey (SNIS)\footnote{SNIS: National system of information on sanitation. Electronic version: app.cidades. gov.br/serieHistorica/}.
The present work will be restricted to two urban metrics, one for each scaling regime: i)\textit{GDP}, a  socio-economic variable that presents a superlinear behavior typically with the population size and ii)\textit{water supply network length}, an infrastructure variable which has a sublinear behavior typically.

A recent work  \cite{meirelles2018evolution} has shown that over 60 variables for the Brazilian urban system are  well described by a power-law equation of the form:
\begin{equation}\label{eq_power_law}
Y_i(t) = Y_0(t) N_i(t)^{\beta_T}. 
\end{equation}
Here, the time-dependent variables $Y_i(t)$ and $N_i(t)$ are relative to the city $i$; the former represents some urban metric (for instance GDP or water network length) and the latter represents the city population size.
The two parameters in Eq.~(\ref{eq_power_law}) are the intercept parameter $Y_0(t)$ and the transversal scaling exponent $\beta_T$, which are obtained by the fit of this power law with the urban system data. These two parameters have to do with the \textit{macro-scale} properties of the urban system and, at first,   do not represent the particularities of a single city - the \textit{micro-scale}. 
As we will show in the next sections, the intercept parameter is a time-dependent variable, while the transversal scaling exponent can or cannot be time-dependent.

\subsection*{Transversal Scaling}

\begin{figure} 
    \begin{center}
        \includegraphics[width=\columnwidth]{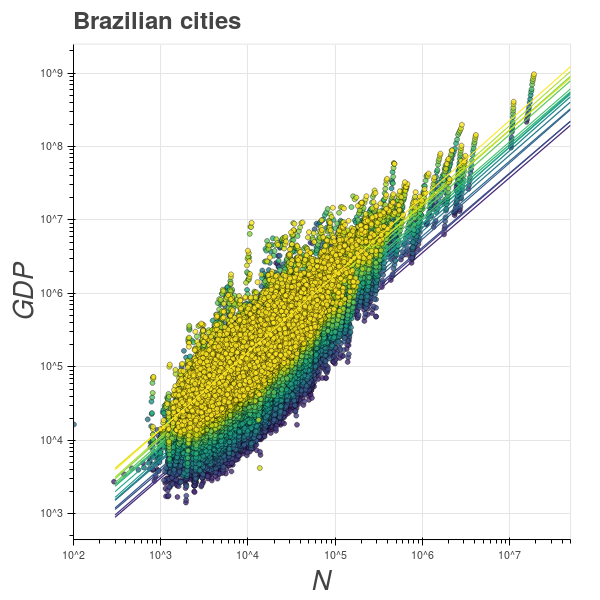}
        \caption{\label{fig_gdpxN_brazil} 
            Scaling relation between population and GDP for Brazilian municipalities, from 1998 (blue) to 2014 (yellow). The straight lines are the best power-law equation fits for each year (by the maximum likelihood method). The straight lines are virtually parallel, which shows that the transversal scaling exponent is constant and robust.
            The scaling exponent is always greater than 1 for all years,         
            with a mean $ \bar{\beta}_T = 1.04$.
            It reveals a superlinear scaling property, compatible with the fact that the GDP is a socio-economic urban variable.
            The numeric time evolution of the transversal scaling exponent and the intercept parameter are shown in Fig.~\ref{Fig_beta_C_xt}. 
        }
    \end{center}
\end{figure}

\begin{figure*} 
    \begin{center}
        a)
        \includegraphics[scale=0.25]{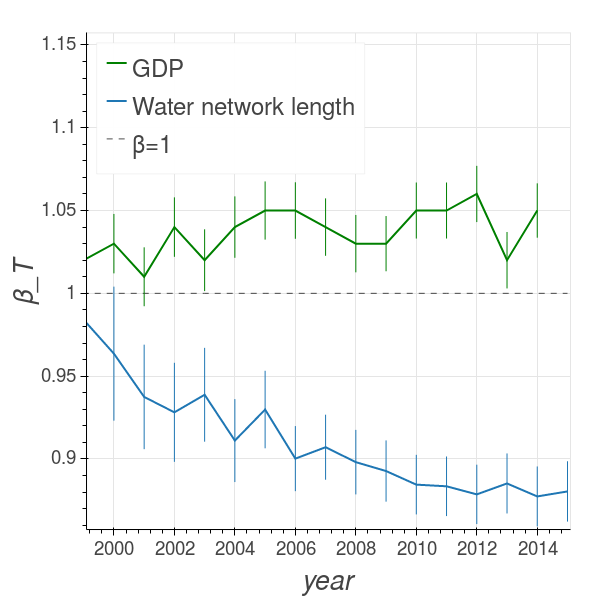}
        b)
        \includegraphics[scale=0.25]{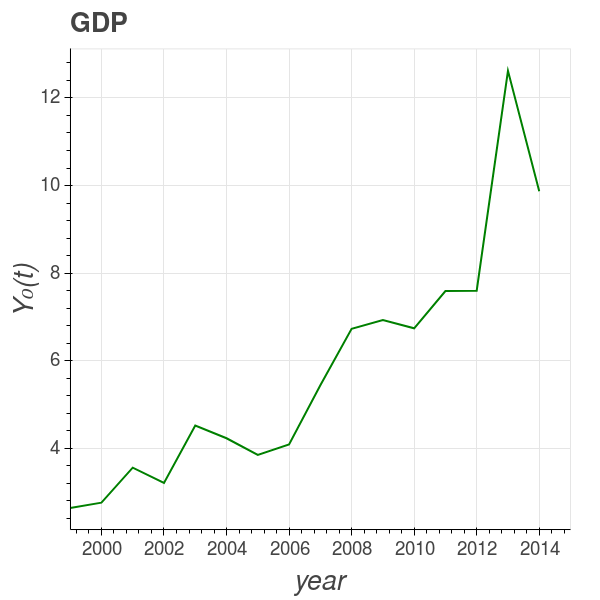}
        c)
        \includegraphics[scale=0.25]{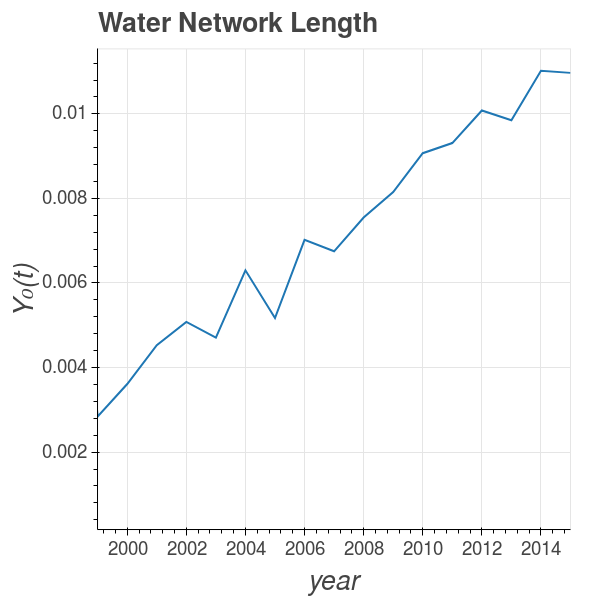}\\
%
\caption{\label{Fig_beta_C_xt}
a) Time evolution of the transversal scaling exponent $\beta_T$ for the GDP (green) and water supply network length (blue) for  Brazilian municipalities. In the GDP case, there is no significant change of this parameter over the years and the regime (superlinear) is always sustained.
            In the case of water supply network length, $\beta_T$ is always smaller than 1, which is expected, given the infrastructure nature of this metric, and it is decreasing over time.
            b) and c)  present the time evolution of the intercept parameter $Y_0(t)$ for GDP and water supply network length, respectively.
            The intercept parameter is constantly growing for both urban metrics.         
        }
    \end{center}
\end{figure*}

Fig.~(\ref{fig_gdpxN_brazil}) shows the GDP as a function of the population size for different years (from 1998 to 2014) of our Brazilian municipality subset.
The straight lines in Fig.~(\ref{fig_gdpxN_brazil}) are the best fit (by the maximum likelihood method) of the Eq.~(\ref{eq_power_law})  for different years. 
The transversal scaling exponent $\beta_T$ (the slope) of each line in Fig.~(\ref{fig_gdpxN_brazil}) is always greater than one, indicating a persistent superlinear behavior. Moreover, the best fit lines are visually parallel, that is,  $\beta_T$  is approximately constant, even with the time evolution of studied municipalities, which reveals the robustness of the scaling exponent.
These facts can be observed in more detail in Fig.~(\ref{Fig_beta_C_xt}-a), which presents the time evolution of $\beta_T$ for the Brazilian municipality subset. The transversal scaling exponent stays approximately constant even with the intercept parameter $Y_0(t)$ continuously increasing with time (see Fig.~(\ref{Fig_beta_C_xt}-b)).

Fig.~(\ref{Fig_beta_C_xt}) also presents the time evolution of the transversal scaling exponent $\beta_T$ for the water supply network length (in blue). In this case, $\beta_T$ is not constant and decreases over time, as seen in Fig.~(\ref{Fig_beta_C_xt}-a) while remaining smaller than 1, which was expected given it refers to an infrastructure variable. 
According to the available data, it is hard to establish whether its value will stabilize or not. The fact that this variable is not constant could suggest that the urban system is still out of balance with respect to this urban metric, as suggested by Pumain's theory \cite{pumain2006}. 
Moreover,  the data suggest that the intercept parameter $Y_0(t)$ of this urban metric, as it was observed in GDP,  maintains a continuous growth through the observed time frame (see Fig.~(\ref{Fig_beta_C_xt}-c)).

\subsection*{Longitudinal Scaling}

\begin{figure*} 
    \begin{center}
        a)\includegraphics[scale=0.27]{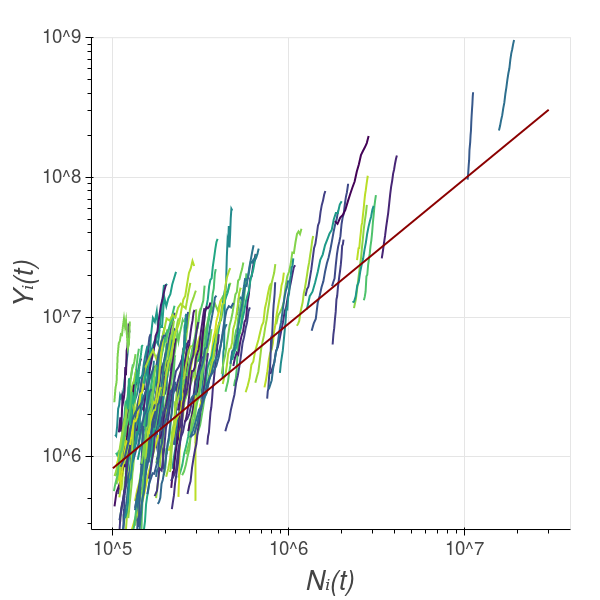}
        b)\includegraphics[scale=0.27]{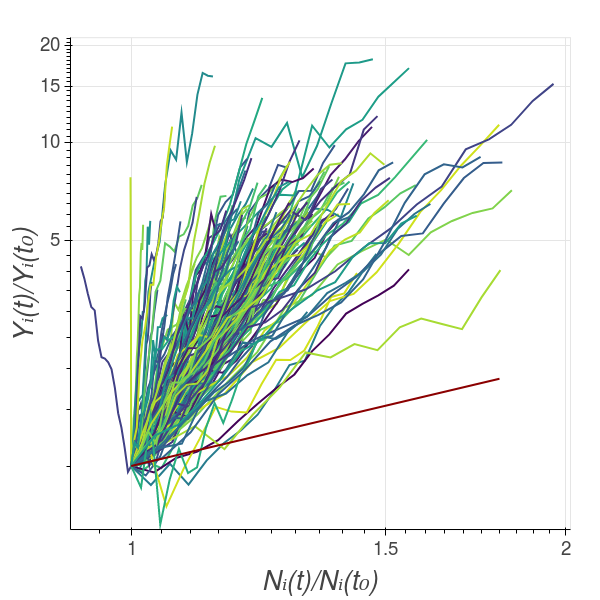}
        c)\includegraphics[scale=0.27]{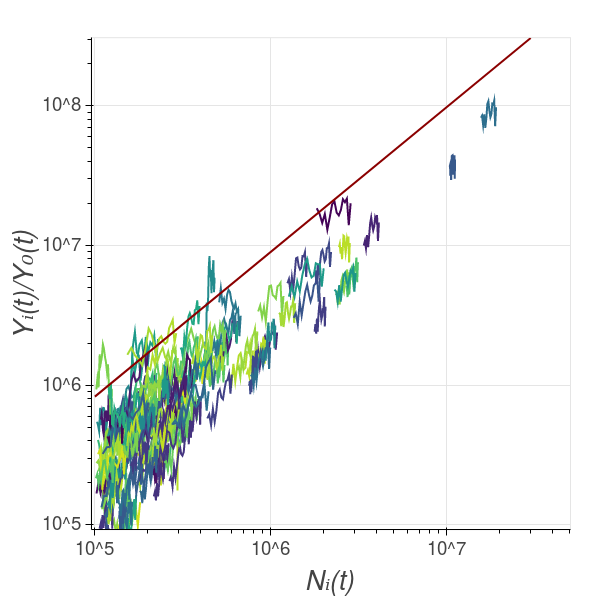}
        d)\includegraphics[scale=0.27]{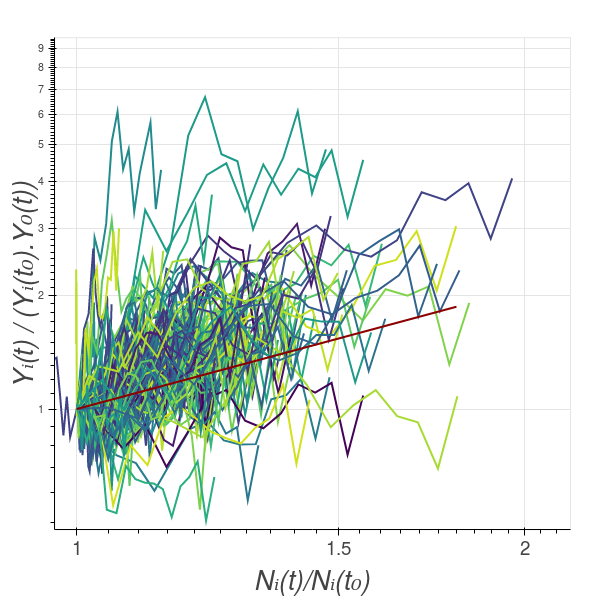}
\caption{\label{Fig_YxN_bigcities}
Different forms to see the longitudinal dynamics of the GDP and the population size for all Brazilian municipality subsets.
Each trajectory represents the time evolution of one single urban area,  from the year 1998 to 2014. %
a) log-log plot of the time evolution of the raw data of GDP as a function of the population size.
The dark red straight line is the power-law equation, with the average transversal scaling exponent $\bar{\beta}_T = 1.05$. 
b) log-log plot of the re-scaled form of the longitudinal dynamics,  which allows us to compare the slopes of the cities' trajectory. This graph shows us that cities have different slopes, and they are greater than  $\bar{\beta}_T$, represented by the dark red line.
c) Decomposed longitudinal trajectory, which allows seeing the dynamics without global effects. d) Decomposed and re-scaled form of the longitudinal dynamics, which shows that the individual slopes are compatible with the transversal scaling exponent, represented by the dark red line.
The distribution of the individual slopes (for raw and decomposed data) can be seen in Fig.~(\ref{fig_hist}).}
    \end{center}
\end{figure*}

We now focus on the individual evolution of Brazilian municipalities. 
Fig.~(\ref{Fig_YxN_bigcities}) presents different ways of observing the longitudinal dynamics of the GDP and the population size for the municipalities subset. 
Fig.~(\ref{Fig_YxN_bigcities}-a) presents the raw longitudinal trajectories, while 
Fig.~(\ref{Fig_YxN_bigcities}-b) presents them re-scaled as $Y_i(t)/Y_i(t_0)$, as a function of $N_i(t)/N_i(t_0)$, 
following the idea proposed in \cite{Barthelemy2017}.  
The re-scaled form allows us to compare in one single image the slopes of the municipalities' trajectories.
Here, $t_0$ is the first year that the data are available. One can see that municipalities experience different slopes, and in all cases, the exponent is greater than the transversal one (given by $\beta_T$ and represented by the dark red line in 
Fig.~(\ref{Fig_YxN_bigcities}-b)).
Similar evidence was reported recently by Depersin and Barthelemy \cite{Barthelemy2017}, which analyzed the temporal dynamics of delay in traffic congestion in US cities. They observed as we did here, that the individual dynamics do not collapse in a single and universal curve, suggesting that longitudinal scaling in cities is not governed by a single universal scaling exponent as the global system is.

Individual municipalities are being pushed by the growth of the global intercept parameter $Y_i(t)$ and will rise in the $\ln Y-$x$-\ln N$ plane, having higher slopes than the global one. One way to deal with this is to \textit{decompose} the longitudinal trajectory, graphing not $\ln Y_i (t)$ in the ordinate, but instead. 
  $\ln Y_i(t)- \ln Y_0 (t)$, that is $\ln(Y_i(t)/Y_0(t))$, in order to eliminate global effects, as suggested by \cite{Hong2018}.
The decomposed longitudinal trajectory is shown in   Fig.~(\ref{Fig_YxN_bigcities}-c), and its re-scaled form is presented in Fig.~(\ref{Fig_YxN_bigcities}-d).
The slopes observed on the decomposed and re-scaled form of the longitudinal trajectories are compatible with the transversal slope, represented by the dark red line in Fig.~(\ref{Fig_YxN_bigcities}-d).

\begin{figure*}[htb!]
    \begin{center}
        \includegraphics[scale=0.3]{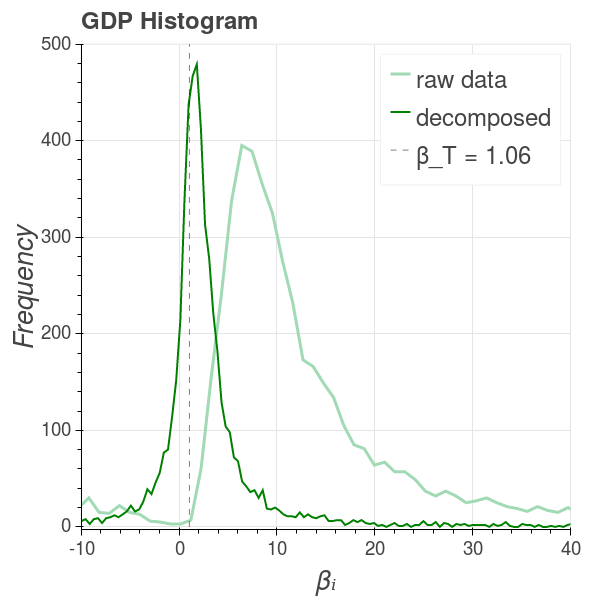}
        \includegraphics[scale=0.3]{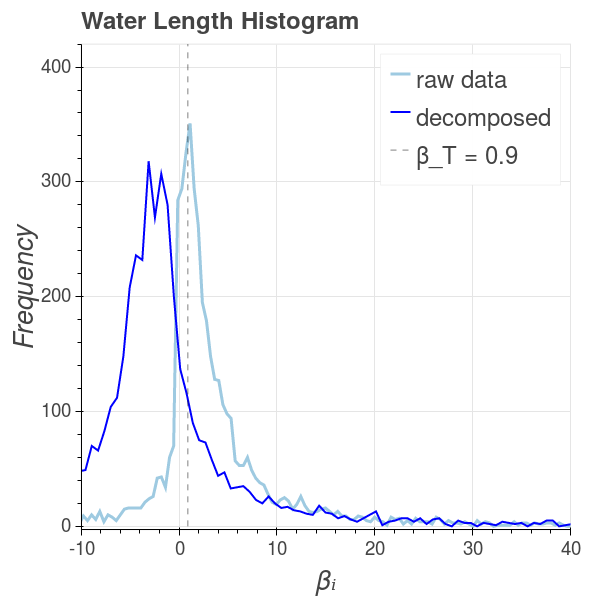}
\caption{\label{fig_hist} Histogram of the longitudinal scaling exponent sets $\{ \beta_i \}_i$ (raw data) and $\{ \beta_i^{dec} \}_i$ (decomposed data) for GDP and water network length for all Brazilian municipality subsets. 
For GDP (on the left), the decomposed data are distributed around the transversal scaling exponent $\beta_T$ (vertical dashed line), suggesting that it makes sense to decompose this urban variable. However,  in the case of water network length (on the right), the distribution of the raw (non-decomposed) data is closer to the global slope than the distribution of the decomposed one, suggesting that decomposition is not working for this urban variable.
}
    \end{center}
\end{figure*}

Lets call $\beta_i$ the scaling exponent of the $i$-th city, that is, the slope of the (raw) trajectories described in Fig.~(\ref{Fig_YxN_bigcities}a-b)  calculated using the longitudinal evolution of $Y_i(t)$ with $N_i(t)$.
Similarly, we can compute the individual decomposed scaling exponent, say $\beta_i^{dec}$, computed using the decomposed longitudinal trajectory described in Fig.~(\ref{Fig_YxN_bigcities}c-d).
Fig.~(\ref{fig_hist}) presents the distribution of the individual slope, for both sets $\{\beta_i\}_i$  and $\{\beta_i^{dec}\}_i$,  for GDP and water supply network length, for all studied municipalities. 
One can see that the decomposed individual slopes for GDP are distributed around the global slope, suggesting that the decomposed version of individual trajectories recover the transversal phenomena for GDP in Brazilian municipalities.
Moreover, it suggests that regardless of each municipality having different dynamics, that is, different longitudinal scaling exponents $\beta_i^{dec}$, their distribution presents a mean value compatible with the transversal scaling exponent. 

However, in the case of the water supply network length, the average of the distribution of the non-decomposed data is closer to the transversal slope than the decomposed one, suggesting that decomposition don't recover the transversal scaling exponent for every urban variable. It is possible that this is the case because the transversal scaling exponent $\beta_T$ for water network length is not stable across the studied years.
These results suggest that the decomposition alone is not enough to infer that the individual and the global systems follow the same scaling properties in every case. In the next sections, we will introduce a theoretical approach that suggests that, in order to have an agreement between transversal and longitudinal scaling, it is also necessary to consider a new ingredient: the city growth rate.


\section{Theoretical Approach }\label{sec_theoretical}

\begin{figure} 
    \begin{center}
        \includegraphics[width=\columnwidth]{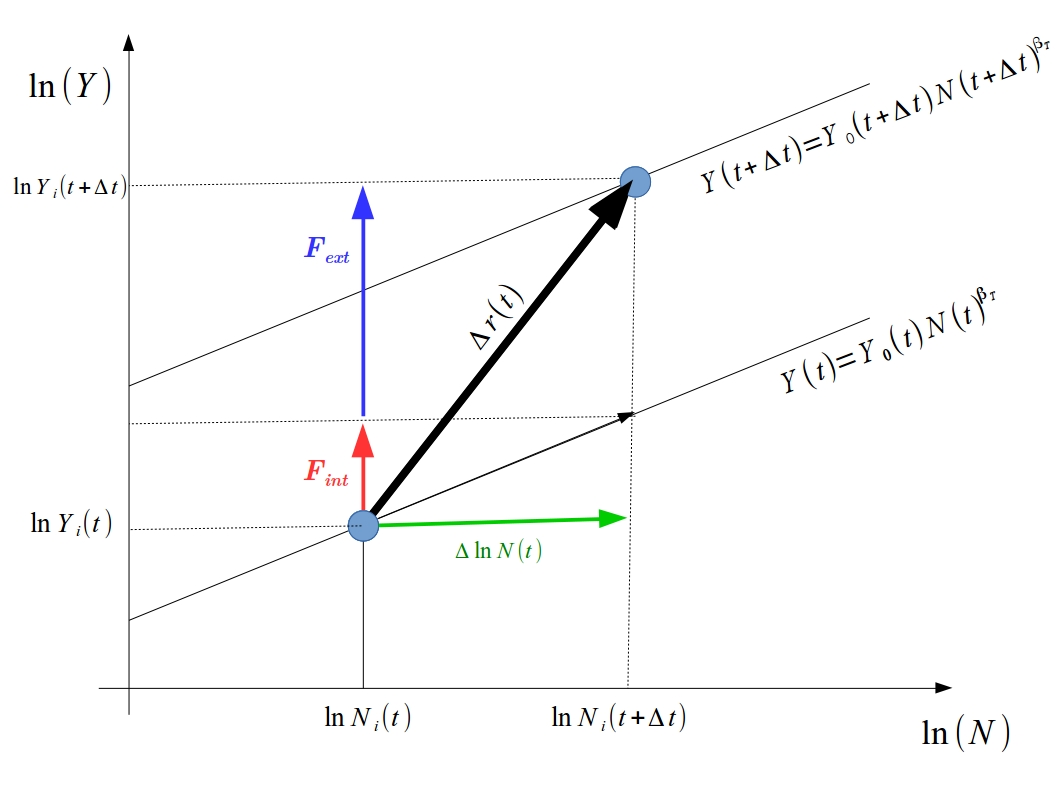}
        \caption{\label{Fig_plane_vector}
            Plane $\ln Y$-\rm{x}-$\ln N$, which represents the ``movement''  of the city as a particle in a vector field. 
            In the horizontal direction, we have the vector (green) that represents the increase in population size. In the vertical direction there is the action of two vectors: $F_{int}$ (red), which  is an extensive quantity whose magnitude is a direct response to the increment of the population size related to an agglomeration effect between the individuals that live in this single city; and $F_{ext}$ (blue), which  is the vector related to some external aspects, or the interaction between the individuals from this city with individuals of other cities, or some individual incorporated ability. The action of this vector field during a time interval $\Delta t$ conducts to a ``displacement'' $\Delta r(t)$ of this city (or particle) in this two-dimensional plane.
            The two parallel lines are given by the global system (transversal) power law~(Eq.~(\ref{eq_power_law})) in $t$ and $t+\Delta t$. 
        }
    \end{center}
\end{figure}

In this section, we present a theoretical approach to describe urban metrics dynamics. In order to do so, we will treat the dynamics of the urban metrics as an analogous problem of particles in a vector field. 
Fig.~(\ref{Fig_plane_vector}) presents the plane $\ln Y$-\rm{x}-$\ln N$ and the two-dimensional ``movement'' of one single city -- a ``particle'' -- as a result of the \textit{vectors} acting in the horizontal or vertical direction. 

In the horizontal direction there is a vector representing an increase in the city's population size. It is coloured in green in Fig.~(\ref{Fig_plane_vector}) and has a magnitude $\Delta \ln N_i (t)$, where we have introduced the compacted notation:
\begin{equation}\label{eq_def_compactN}
\Delta \ln N_i (t) \equiv  \ln N_i (t+ \Delta t) - \ln N_i(t), 
\end{equation}
as it was suggested in \cite{Hong2018}.
In the vertical direction of this plane 
we have vectors acting on the increment of the urban metric (GDP or water supply network length, for instance).
We will consider, by hypothesis, that there are at least two distinct vectors acting in this direction.
The first, let's say $F_{int}$, represented by the red vector in Fig.~(\ref{Fig_plane_vector}),  is an
\textit{extensive} quantity 
Whose magnitude is a direct response to the increase in population size. This vector has to do with the \textit{agglomeration effect} that came from the interaction between the individuals who belong to this single city. 
The second, let's say $F_{ext}$, represented by the blue  vector in  Fig.~(\ref{Fig_plane_vector}), is the result of all external mechanisms such as, for instance, some wealth/knowledge that comes from other cities or regions;  it can also represent the result of the interaction 
between individuals who belong to this single city with dwellers from other cities; or even some individual incorporated ability that increases the individual productivity.

Therefore, the resulting vector acting on the vertical direction of the plane, say $F_{tot}$, is the sum of these two vectors, that is:
\begin{equation}\label{Eq_Ftot}
F_{tot}  = F_{int}  + F_{ext},
\end{equation}
which has a magnitude 
\begin{equation}\label{eq_def_compactY}
F_{tot}  =  \Delta \ln Y_i(t) \equiv  \ln Y_i (t+ \Delta t) - \ln Y_i(t).
\end{equation}

The action of this vector field (in the horizontal and vertical directions) during a time interval $\Delta t$ conducts to a ``displacement''  $\Delta r(t)$ of this city (or particle) in the two-dimensional plane $\ln Y$-\rm{x}-$\ln N$. 
But let us try to identify these vectors with the empirical variables available.  

The data presented in the previous section suggest that we have an empirical law that governs the cities, which can be described by the expression~(\ref{eq_power_law}). If this equation is a law, then any theory that is formulated to describe scaling properties in cities must be constrained to follow it. 
As this equation holds for any time $t$, we can write it for the next time instant $t+\Delta t$, that is:
\begin{equation}\label{eq_empirical_law2}
Y_i(t + \Delta t) = Y_0(t+ \Delta t)N_i(t+ \Delta t)^{\beta_T(t+ \Delta t)}.
\end{equation}

Then, by extracting the logarithm of the ratio $Y_i(t + \Delta t)/Y_i(t)$ and using Eqs.~(\ref{eq_power_law}) and~(\ref{eq_empirical_law2}) we are conducted to:
\begin{equation}\label{eq_deltaY}
\Delta \log Y_i(t) = \log \left( \frac{Y_0(t + \Delta t)}{Y_0(t)} \right) +  (\bar{\beta}_T  + \epsilon) \Delta \log N_i(t).
\end{equation}
where we used the compacted forms defined on~(\ref{eq_def_compactN}) and~(\ref{eq_def_compactY}). 
Moreover, we also introduced $\bar{\beta}_T$ as the average value of the transversal exponent during the time interval $\Delta t$, and the parameter $\epsilon$, which is a quantity proportional to the difference  $\beta_T(t+\Delta t) - \beta_T(t)$. In fact, the data analysis suggests that $\epsilon$ is sufficiently small for the cases we are studying here, so it will be neglected in our analyses. When $\beta_T(t)$ is constant, as it is approximately the case for GDP dynamics, then $\epsilon = 0$.

The elements of  Eq.~(\ref{eq_deltaY}) can be identified with the vectors presented in Fig.~(\ref{Fig_plane_vector}) and consequently with  Eq.~(\ref{Eq_Ftot}). It allows us to identify:
\begin{eqnarray}\label{eq_Fext_teo}
F_{ext} = \log \left( \frac{Y_0(t + \Delta t)}{Y_0(t)} \right)
\end{eqnarray}
and 
\begin{eqnarray}
F_{int} = (\bar{\beta}_T + \epsilon) \Delta \log N(t).
\end{eqnarray}

The external vector, since it is directly computed from the ratio between the final and initial intercept parameter,  can be interpreted as a measurement of the global growth of the urban metric. In this sense, the value given by~(\ref{eq_Fext_teo}) is an average value of the external vector.  That is: typically, a city in the system has an external vector magnitude given by the value computed from~(\ref{eq_Fext_teo}).
In the last section, when we decomposed each city's evolution into a relative change, we removed external factors acting on each city and considering only internal factors (the ones that come from agglomeration effects).
In relation to the magnitude of the internal vector, it is an extensive variable; that is,  it is a direct response to the increase of the population size. 
These results suggest that, in order for the urban metric to depend only on the population size (under the form  $Y = \textrm{cte} \cdot N^{\beta_T}$), 
it is necessary for $\beta_T$ to be constant ($\epsilon \to 0$) and $F_{ext} \to 0$, which means absence of global growth.  That can be the case for some urban metrics, but of course, it is not the case for GDP and many other variables. Our theoretical approach suggest that $\overline{\beta_i^{dec}} \neq \beta_T$ when $\epsilon \neq 0$, which was observed in our empirical data for the water supply network length.

\subsection*{Relation between transversal and longitudinal scaling exponents}

\begin{figure*} 
    \begin{center}
        \includegraphics[scale=0.35]{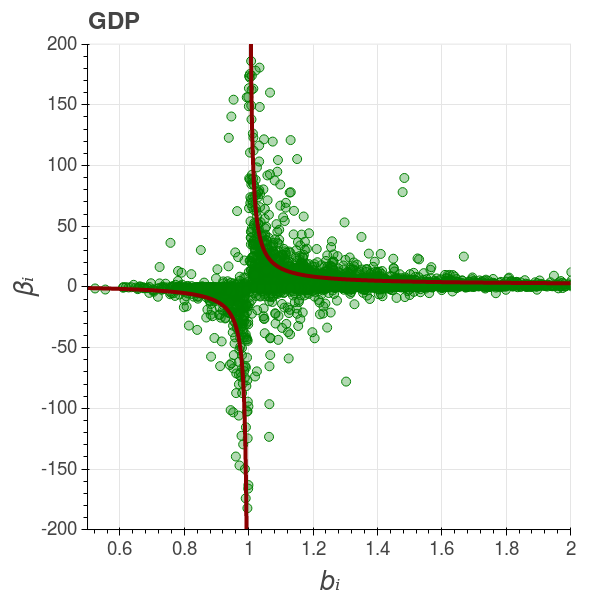}
        \includegraphics[scale=0.35]{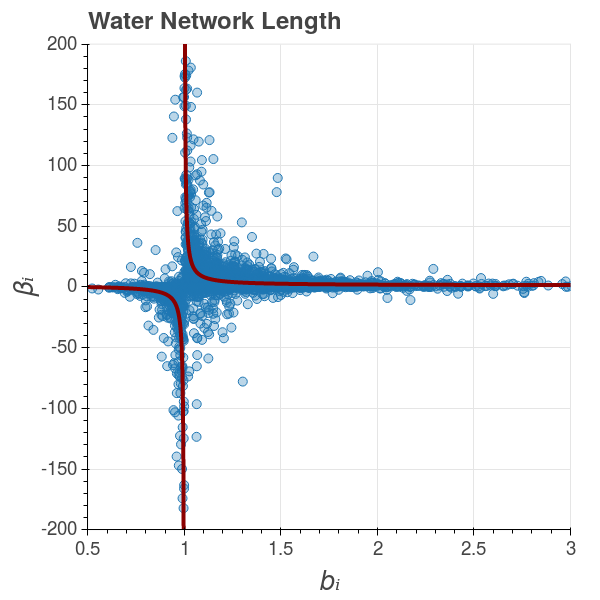}
        \caption{\label{Fig_betaixbi} Graph of $\beta_i$ as a function of the population growth rate $b_i$, for both GDP (on the left) and water network length (on the right).  Each  dot represents the data of a single Brazilian municipality subset and the red curve is the theoretical prediction given by Eq.~(\ref{eq_betai_bi}) using:  $\epsilon =0$ for both cases;  $\beta_T = 1.15$ for GDP;  and $\beta_T = 0.9$ for water network length. This result illustrates the strong dependence between  the longitudinal scaling exponent and the population growth rate of the city. It also suggests that  cities with bigger $\beta_i$ are the ones with  little or no growth ($b_i \approx 1$). 
            Moreover, $b_i < 1$ (decreasing population) implies a negative $\beta_i$.}    
    \end{center}
\end{figure*}

With the approach presented above, it is possible to write a relation between the transversal and the longitudinal scaling exponent.
Given that the longitudinal scaling exponent $\beta_i$ is obtained by:
\begin{equation}\label{eq_betai}
\beta_i = \frac{ \Delta \ln Y_i(t)}{\Delta \ln N_i(t)},
\end{equation} 
then if we divide  Eq.~(\ref{eq_deltaY}) by $\Delta \ln N_i(t)$,  we have:
\begin{equation}\label{eq_betai_bi} 
\beta_i = \beta_T + \epsilon +  \frac{F_{ext}}{\ln b_i},
\end{equation}
where  $b_i \equiv N_i(t+ \Delta t) / N_i(t)$ is the city population growth  rate. 
The graphs in Fig.~(\ref{Fig_betaixbi})  show that this result works very well when we analyze $\beta_i$ as a function of $b_i$, 
for both GDP and water supply network length for the studied municipalities. It shows a strong dependence between these two variables. 

The result~(\ref{eq_betai_bi}) also suggests that if $F_{ext} > 0$ and $b_i>1$, which means that both the intercept parameter (global growth)  and the population are growing with time, then $\beta_i$  will always be greater than the global exponent $\beta_T$. 
The increment in the intercept implies a more accentuated slope of the city trajectory in the plane $\ln Y$-x-$\ln N$ (that is, bigger $\beta_i$) in relation to the transversal trajectory (related to $\beta_T$), in accordance with the empirical observation presented in this study as well as other evidences presented in recent literature \cite{Barthelemy2017,Hong2018,Zhao2018}. 
Moreover, longitudinal and transversal scaling will be the same when only internal factors (agglomeration effects) are acting on the system ($F_{ext}=0$). 

In urban scaling analysis, it is important to know the value of the scaling exponent because it gives us the efficiency and productivity of the city or the urban system (given by $\beta_i$ and $\beta_T$, respectively) since it shows how the urban metric reacts to an increase in population. 
For instance, in socio-economic variables, larger values of $\beta$ mean a more productive city, and for infrastructure variable, smaller values means a more efficient city.
However, in the context we are analyzing, cities with very large values of  $\beta_i$ are not necessarily more productive.
In fact, large values of scaling exponents are related to cities with a very low growth rate
(according to Eq.~(\ref{eq_betai})),  so the value of $\beta_i$ is not useful for the main purpose of seeing how productivity or scaling economies emerge.

\begin{figure*} 
    \begin{center}
        A)\includegraphics[scale=0.3]{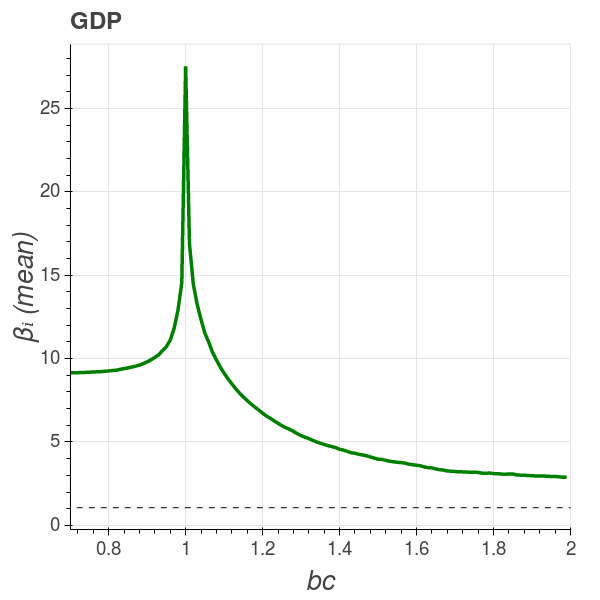}
        B)\includegraphics[scale=0.3]{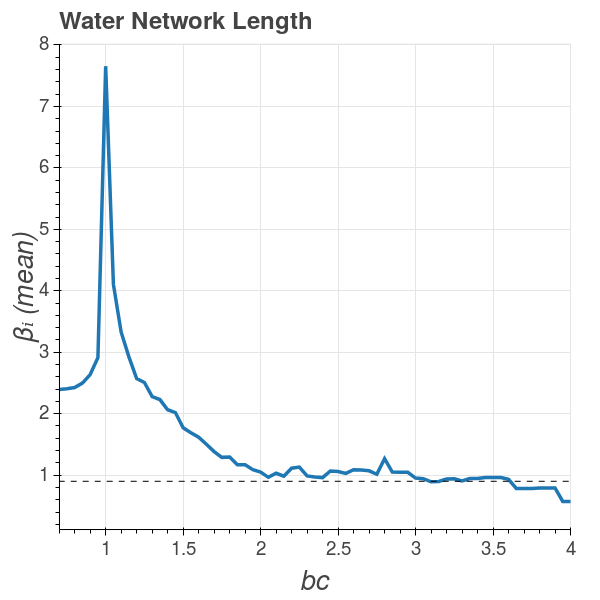}
        C)\includegraphics[scale=0.3]{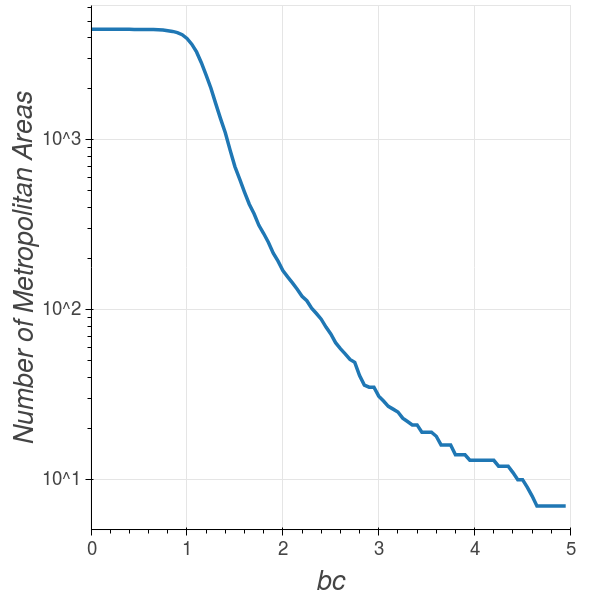}
        \caption{\label{Fig_bcorte}
            Graph A) and B):  mean value of the longitudinal scaling exponent (for GDP and water network length) as a function of the growth rate threshold $b_c$. The parameter $b_c$ delimits the cities that will be used to compute the average.
            The mean value of $\bar{\beta_i}$ decreases drastically as $b_c$ increases. Moreover 
            the greater  $b_c$ is the more $\bar{\beta_i}$ approaches to $\beta_T$ (represented by the dashed line).
            It's reasonable to think that in an ideal situation with no external forces and with a significant number of cities with the larger growth rate for better statistics, the average of the longitudinal scaling exponent will converge to the value of the transversal scaling exponent. Graph C: number of metropolitan areas used to compute the mean as a function of $b_c$. This number is drastically smaller for greater $b_c$ values.
        }    
    \end{center}
\end{figure*}

However, we believe that the value of $\beta_i$ will informs about the city's efficiency when it has a sufficiently large population growth rate.  In order to investigate this, we computed the average values of the longitudinal scaling exponents $\bar{\beta_i}$, using  only cities with $b_i$ greater than a threshold $b_c$, and later built the graph presented in  Fig.~(\ref{Fig_bcorte}), where we can see that $\bar{\beta_i}$ decreases drastically for greater values of $b_c$, approaching the transversal exponent value for both GDP and water network length.    
This result suggests that when we consider a city that has grown significantly during the time period analyzed, it is relevant to understand its longitudinal scaling growth properties as a microscopic version of the macroscopic growth of the urban system. 
Another important aspect of these findings is that only \textit{decomposition} is not enough to link globally with longitudinal scaling, as highlighted in the last section. In fact, decomposition only makes sense if we consider cities with sufficient growth in a given period of time, or with constant transversal scaling exponent $\beta_T$ ($\epsilon \neq 0$).

The results presented here must be confronted with more urban metrics and other countries. Moreover, a problem resulting from the approach presented in this section concerns the small number of municipalities that present $b_c$ sufficiently large. For instance, in order to compute the average $\bar{\beta_i}$ for $b_c > 4$ we used only 13 municipality subsets (see Fig.~(\ref{Fig_bcorte}-c)).
The statistics could be improved if we were studying an urban system with more municipalities experiencing higher growth rates, but maybe such systems don't even exist.
Thus a more feasible situation for future analyses consists of finding a way to normalize the longitudinal scaling exponent to the city's growth rate.



%

\subsection*{The external Vector}

Eq.~(\ref{eq_Fext_teo}) represents the average magnitude of the external vector, that is, a city within the system will have an external vector with magnitude typically given by this value.  
However, it is interesting to know about the specific external vector value acting on an individual city. This is a very difficult matter to be resolved given that it involves particularities of the city, but we can infer its answer by the data that we have available. 

\begin{figure*} 
    \begin{center}
        \includegraphics[scale=0.35]{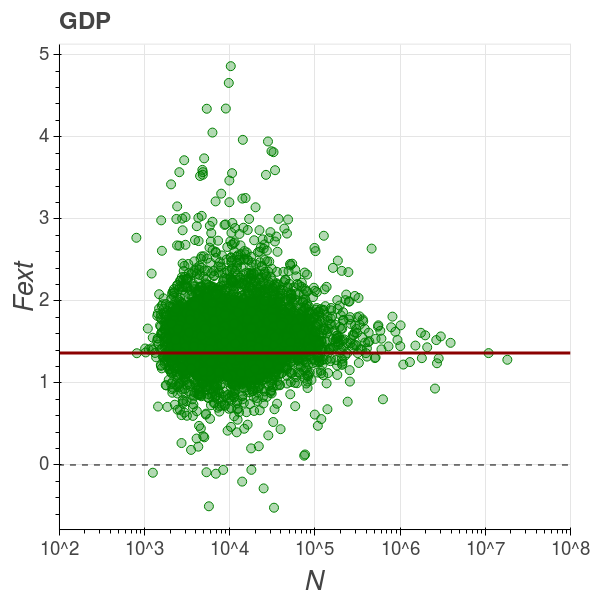}
        \includegraphics[scale=0.35]{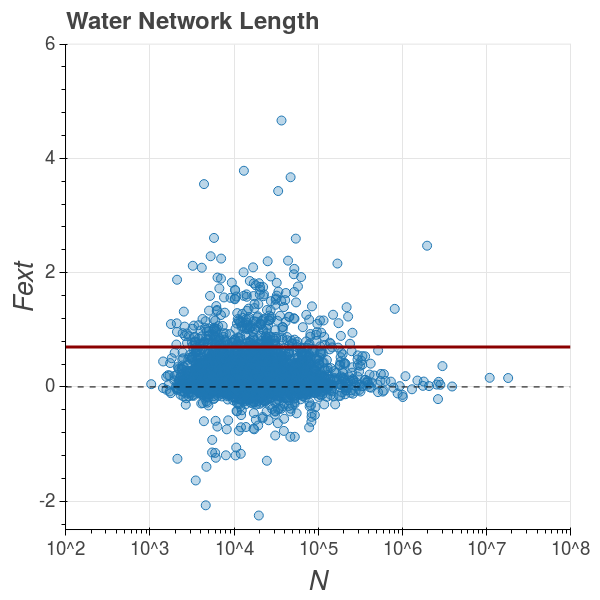}
        \caption{\label{Fig_Fext}
            Magnitude of the external vector as a function of population size for every single Brazilian municipality in our subset.
            On the left, we have data referring to GDP and on the right to water network length. The dots represent $F_{ext}^i$ computed from the expression~(\ref{eq_Eext_i}) while the red line represents the average magnitude of this vector over the system, 
            computed by the expression~(\ref{eq_Fext_teo}).
            The dashed line is $F_{ext}=0$.
            In the case of GDP, the municipality subsets of all sizes are distributed around the average value, but in the case of the water network length, the greater subsets presented external vector smaller than the average. These particularities imply different dynamics of the transversal scaling exponent, as shown in Fig.~(\ref{Fig_Fext_esquema}).  }
    \end{center}
\end{figure*}

For instance, we can use the result given by  Eq.~(\ref{eq_betai_bi}) to infer the external vector  $F_{ext}^i$ relative to the $i$-th city. That is, we can write that:
\begin{equation}\label{eq_Eext_i} 
F_{ext}^i =  (\beta_i -\beta_T - \epsilon) \ln b_i,
\end{equation}
and if  $\beta_T$,  $\beta_i$ (given by Eq.~(\ref{eq_betai})) and $b_i$
are known, then it is possible to estimate (assuming $\epsilon \approx 0$) the  individual  external vector. That is the case presented by Fig.~(\ref{Fig_Fext}), where each dot represents the value obtained for the external vector of a single municipality, for both GDP and water network. 

\begin{figure*} 
    \begin{center}
        \includegraphics[scale=0.3]{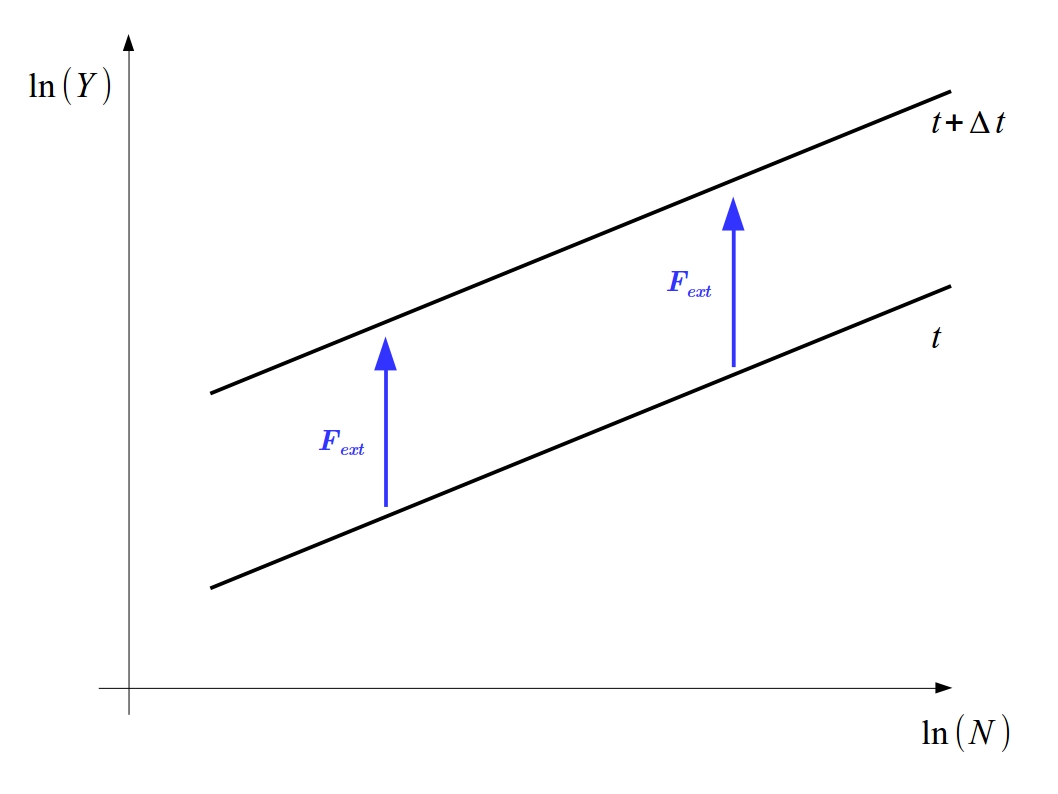}
        \includegraphics[scale=0.3]{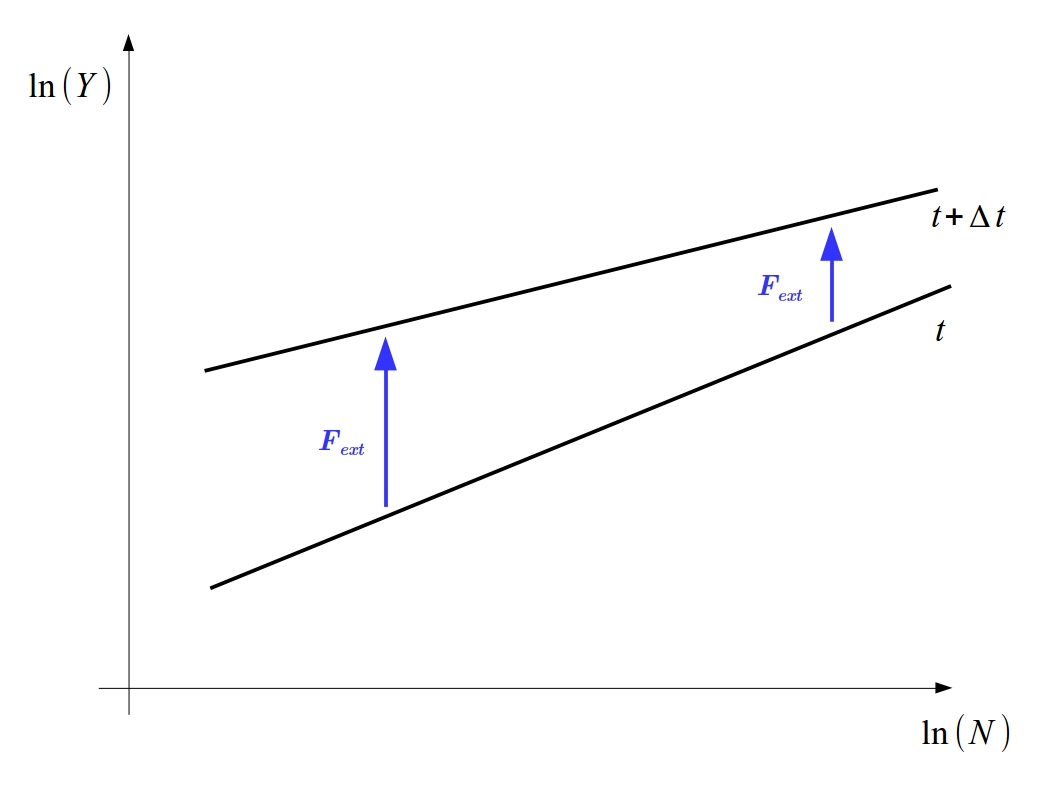}
        \caption{\label{Fig_Fext_esquema}
            Schematic drawing representing the plane
            $\ln Y$-\rm{x}-$\ln N$ with two scenarios for the external vectors. In the first scenario (on the right), the magnitude of the external vectors is the same regardless of city size; implying that the slope of the fit line (in
            $\ln Y$-\rm{x}-$\ln N$ plane) remains constant from $t$ to $t+\Delta t$. That is more or less what happens in the GDP of the Brazilian municipalities, revealing that this urban variable is in a mature state in the system. In the second scenario (on the left), the external vector is smaller for bigger cities, which implies that the slope of the fit line decreases with time (from $t$ to $\Delta t$). That is apparently the case for the water network length of the Brazilian municipalities, suggesting that this urban metric is not mature in the system.  }    
    \end{center}
\end{figure*}

Fig.~(\ref{Fig_Fext}) also presents the comparison between this individual and the average external vector magnitudes. 
It suggests an interesting aspect differentiating these two urban metrics' dynamics.
In the case of GDP, the cities of all sizes are distributed around the average magnitude of the external vector. 
However, in the case of water supply network length, bigger cities present external vector smaller than the average. These characteristics imply different dynamics with respect to the transversal exponent, according to the schematic drawing in  Fig.~(\ref{Fig_Fext_esquema}), which presents the plane
$\ln Y$-\rm{x}-$\ln N$ with two scenarios for the external vectors.
In the first scenario,  the external vector is approximately the same for all cities of the system, regardless of their size; it implies that the slope of the fit line (in $\ln Y$-\rm{x}-$\ln N$ plane) remains constant. 
That is more or less what happens in the GDP context of the Brazilian municipalities.
It suggests an equilibrium situation, or at least that this urban variable is in a mature state inside the system. 

In the second scenario, the external vector is smaller for bigger cities, which implies that the slope of the fit line decreases over time. That is apparently the case for the water network of the Brazilian municipalities.  
One possible explanation is that the system is still out of equilibrium. That is,  water network length in Brazil is not mature yet, and maybe it will converge to the equilibrium (when the magnitude of the external vector of all cities will be around an average value) given enough time. 
In any case, in order to have a better understanding of that aspect, further research about these observations is necessary, which can be achieved by following the evolution of more urban variables.

\section{Conclusion}\label{sec_conclusion}

We analyzed the longitudinal and transversal scaling dynamics of 5507 Brazilian municipalities, aggregated with contiguous dense surrounding municipalities from the totality of municipalities. 
We showed, using two urban metrics - GDP and water supply network length - 
That the longitudinal scaling exponents are different from each other, but they are distributed around an average that approaches the transversal scaling exponent when we remove external factors (by decomposition) and when we consider cities that grew sufficiently during the analyzed period. We then proposed a formal vectorial description that describes under which conditions longitudinal ($\beta_i$) and transversal ($\beta_T$) scaling should converge or where we can expect discrepancies. This result supports the hypothesis that longitudinal and transversal urban dynamics could be differently scaled versions of the same phenomenon. However, in order to have a more conclusive argumentation,  further investigation is required with other urban variables and other countries.

\section*{Acknowledgments}

We would like to acknowledge all colleagues from the Mathematical Department of City, University of London, where most of the work was done. FLR acknowledges the members of CASA-UCL, especially the stimulating discussions with Elsa Arcaute during his sabbatical year in 2017. Also, thanks to the Brazilian agencies CAPES
(process number: 88881.119533/2016-01) and CNPq (process number: 405921/2016-0) for financial support. 

\section*{Author Contributions}

FLR, JM, and AB conceived the project; FLR, JM, and AB designed research; FLR and JM curated the data; FLR and JM analyzed the data; FLR, JM, VMN, CRN, and AB analyzed and interpreted the results; FLR and JM elaborated the formal analysis; FLR wrote the original draft; FLR, JM, VMN, CRN, and AB read, and commented on, the manuscript.

\bibliographystyle{ieeetr}
\bibliography{vini.bib,fabiano.bib}

\end{document}